\documentclass[conference]{IEEEtran}
\pdfoutput=1
\usepackage{amsmath,amssymb,amsfonts}
\usepackage{cuted}
\usepackage{array}
\usepackage{textcomp}
\usepackage{stfloats}
\usepackage{url}
\usepackage{verbatim}
\usepackage{graphicx}
\usepackage{subcaption}
\usepackage{titlesec}
\usepackage{graphicx}
\usepackage{caption}
\usepackage{subcaption}
\usepackage{cite}
\usepackage{booktabs}
\usepackage{xcolor}
\usepackage{epsfig}
\usepackage{epstopdf}
\usepackage{acronym}
\graphicspath{{fig/}}
\usepackage{mathtools}
\usepackage{amsthm}
\usepackage{bm}
\usepackage{comment}
 \usepackage{setspace}
  \usepackage{lipsum}
  \usepackage[absolute,overlay]{textpos}
  \usepackage{upgreek}
  \usepackage{siunitx}
  \usepackage[greek,english]{babel}

\usepackage{tikz}
\usetikzlibrary{tikzmark,calc}

\usepackage{ragged2e}
\usepackage{textgreek}

\newacro{ACDD}{Alamouti with cyclic delay diversity}
\newacro{URLLC}{ultra-reliable low-latency communications}
\newacro{3GPP}{third generation partnership project}
\newacro{PHY}{physical layer}
\newacro{MIMO}{multiple-input multiple-output}
\newacro{SIMO}{single-input multiple-output}
\newacro{MISO}{multiple-input single-output}
\newacro{SISO}{single-input single-output}
\newacro{MRC}{maximum-ratio combining}
\newacro{SNR}{signal-to-noise ratio}
\newacro{CP}{cyclic prefix}
\newacro{CDD}{cyclic delay diversity}
\newacro{FSC}{frequency-selective channel}
\newacro{STC}{space-time coding}
\newacro{FFT}{fast Fourier transform}
\newacro{LMMSE}{linear minimum mean-squared error}
\newacro{FER}{frame error rate}
\newacro{OFDM}{orthogonal frequency division multiplexing}
\newacro{OCDM}{orthogonal chirp division multiplexing}
\newacro{FSC}{frequency-selective channel}
\newacro{CSI}{channel state information}
\newacro{LMMSE-PIC}{linear minimum mean squared error with parallel interference cancellation}
\newacro{PFE}{perfect-feedback equalizer}
\newacro{FD}{full-duplex}
\newacro{PDP}{power delay profile}
\newacro{PDF}{probability density function}
\newacro{DFT}{discrete Fourier transform}
\newacro{SDFT}{sparse DFT}
\newacro{ICI}{inter-carrier interference}
\newacro{OTFS}{orthogonal time frequency space}
\newacro{AWGN}{additive white Gaussian noise}
\newacro{SWH}{sparse Walsh-Hadamard}
\newacro{LLR}{log-likelihood ratio}
\newacro{PMF}{probability mass function}
\newacro{CRC}{cyclic redundancy check}
\newacro{PAM}{pulse amplitude modulation}
\newacro{QAM}{quadrature amplitude modulation}
\newacro{FWHT}{fast Walsh-Hadamard transform}
\newacro{MAP}{maximum a-posteriori}
\newacro{SC}{single-carrier}
\newacro{ISI}{inter-symbol interference}
\newacro{ZP}{zero-padding}
\newacro{EVD}{eigenvalue decomposition}
\newacro{BCJR}{Bahl, Cocke, Jelinek, and Raviv}
\newacro{WHT}{Walsh-Hadamard transform}
\newacro{APP}{a-posteriori probability}
\newacro{SILE-EPIC}{self-iterated linear equalizer with expectation propagation}
\newacro{EP}{expectation propagation}
\newacro{i.i.d.}{independent and identically distributed}
\newacro{CWCU}{component wise conditionally unbiased}
\newacro{MSE}{mean squared error}
\newacro{EXIT}{extrinsic information transfer}
\newacro{MI}{mutual information}
\newacro{PAPR}{peak-to-average power ratio}
\newacro{DFT-s}{discrete Fourier transform-spread}
\newacro{AMP}{approximate message passing}
\newacro{GAMP}{generalized \ac{AMP}}
\newacro{VAMP}{vector \ac{AMP}}
\newacro{RSC}{recursive systematic convolutional}
\newacro{QPSK}{quadrature phase-shift keying}
\newacro{CFAR}{constant false alarm rate}
\newacro{PD}{probability of detection}
\newacro{PFA}{probability of false alarm}
\newacro{RV}{random variable}
\newacro{CDF}{cumulative distribution function}
\newacro{HD-ZP}{half-duplex ZP}
\newacro{FD-CP}{full-duplex ZP}
\newacro{DFRC}{dual-function radar communication}
\newacro{SINR}{signal-to-interference noise ratio}
\newacro{ISAC}{integrated sensing and communication}
\newacro{SI}{self-interference}
\newacro{RSI}{residual self-interference}
\newacro{ADC}{analog-to-digital converter}
\newacro{DAC}{digital-to-analog converter}
\newacro{ED}{energy-detection}
\newacro{IDFT}{inverse discrete Fourier Transform}
\newacro{SFFT}{symplectic finite Fourier transform }
\newacro{CRB}{Cram{\'{e}}r-Rao bound}
\newacro{ZC}{Zadoff-Chu}
\newacro{RMSE}{root mean square error}
\newacro{UW}{unique word}
\newacro{GFDM}{generalized frequency division multiplexing}
\newacro{RRC}{root-raised cosine}
\newacro{UB}{upper bound}
\newacro{CEF}{channel estimation field}
\newacro{TRX}{transceiver}
\newacro{IF}{intermediate frequency}
\newacro{RF}{radio frequency}
\newacro{FPGA}{field programmable gate arrays}
\newacro{SDR}{software-defined radio}
\newacro{UWB}{ultra wideband}
\newacro{PCB}{printed circuit board}
\newacro{SMA}{SubMiniature version A}
\newacro{MUSIC}{multiple signal classification}
\newacro{CIR}{channel impulse response}
\newacro{FR}{Frequency Range}
\newacro{mmWave}{millimeter wave}
\newacro{LoS}{line-of-sight}
\newacro{ULA}{unform linear array}
\newacro{AoA}{angle-of-arrival}
\newacro{AoD}{angle-of-departure}
\newacro{FIM}{Fisher Information Matrix}
\newacro{TDoA}{time difference of arrival}
\newacro{LM}{Levenberg-Marquardt}
\newacro{RSSI}{received signal strength indicator}
\newacro{OMP}{orthogonal matching pursuit}
\newacro{NLoS}{non line-of-sight}
\newacro{FDoA}{frequency difference-of-arrival}
\newacro{WSNR}{waterlfall SNR}
\newacro{ML}{maximum-likelihood}
\newacro{GPS}{global positioning system}
\newacro{CG}{column generation}
\newacro{MNO}{Mobile Network Operator}

\def\BibTeX{{\rm B\kern-.05em{\sc i\kern-.025em b}\kern-.08em
		T\kern-.1667em\lower.7ex\hbox{E}\kern-.125emX}}
	
\usepackage{soul,color}

\usepackage{tikz}
\usepackage{pgfplots}
\usetikzlibrary{shapes,arrows}
\usetikzlibrary{positioning,calc}
\usetikzlibrary{decorations.pathreplacing,calligraphy}
\usetikzlibrary{arrows.meta}
\usepackage{siunitx}

\tikzset{add/.style n args={4}{
		minimum width=3mm,
		path picture={
			\draw[black] 
			(path picture bounding box.south east) -- (path picture bounding box.north west)
			(path picture bounding box.south west) -- (path picture bounding box.north east);
			\node at ($(path picture bounding box.south)+(0,0.13)$)     {\tiny #1};
			\node at ($(path picture bounding box.west)+(0.13,0)$)      {\tiny #2};
			\node at ($(path picture bounding box.north)+(0,-0.13)$)        {\tiny #3};
			\node at ($(path picture bounding box.east)+(-0.13,0)$)     {\tiny #4};
		}
	}
}

\tikzset{add2/.style n args={4}{
		minimum width=1mm,
		path picture={
			\draw[black] 
			(path picture bounding box.south) -- (path picture bounding box.north)
			(path picture bounding box.west) -- (path picture bounding box.east);
			\node at ($(path picture bounding box.south)+(0,0.13)$)     {\tiny #1};
			\node at ($(path picture bounding box.west)+(0.13,0)$)      {\tiny #2};
			\node at ($(path picture bounding box.north)+(0,-0.13)$)        {\tiny #3};
			\node at ($(path picture bounding box.east)+(-0.13,0)$)     {\tiny #4};
		}
	}
}

\usepackage{pgfplots}
\usepgfplotslibrary{polar}
\usepackage{tikz}

\counterwithin{corollary}{proposition} 

\definecolor{applegreen}{rgb}{0.55, 0.71, 0.0}
\definecolor{awesome}{rgb}{1.0, 0.13, 0.32}
\definecolor{azure(colorwheel)}{rgb}{0.0, 0.5, 1.0}
\definecolor{darklavender}{rgb}{0.45, 0.31, 0.59}
\definecolor{cyan(process)}{rgb}{0.0, 0.72, 0.92}
\definecolor{brightmaroon}{rgb}{0.76, 0.13, 0.28}
\definecolor{ao(english)}{rgb}{0.0, 0.5, 0.0}
\definecolor{brightturquoise}{rgb}{0.03, 0.91, 0.87}
\definecolor{bondiblue}{rgb}{0.0, 0.58, 0.71}
\definecolor{atomictangerine}{rgb}{1.0, 0.6, 0.4}
\definecolor{classicrose}{rgb}{0.98, 0.8, 0.91}
\definecolor{copperrose}{rgb}{0.6, 0.4, 0.4}

\usepackage{etoolbox}

\usepackage{algorithm}
\usepackage{algpseudocode}

\algnewcommand{\Input}[1]{\State \textbf{Input:} #1}
\algnewcommand{\Output}[1]{\State \textbf{Output:} #1}
\AtBeginDocument{\everymath{\small} \everydisplay{\small}}

\begin{document}

\title{5 Shades of Cooperation:\\Spectrum Sharing in the Upper-Mid Band}




\author{\IEEEauthorblockN{Alberto Ceresoli, Marco Mezzavilla, Ilario Filippini, Antonio Capone}
\IEEEauthorblockA{
\textit{Dipartimento di Elettronica, Informazione e Bioingegneria} \\
\textit{Politecnico di Milano}, Milan, Italy \\
\textit{name}.\textit{surname}@polimi.it}\vspace{-1cm}
}

\maketitle

\vspace{-0.2cm}
	
\maketitle

\begin{abstract}
Spectrum exclusively licensed to each mobile network operator (MNO) is scarce, and assigning it in fixed, frequency-orthogonal blocks leaves much of it idle under heterogeneous, time-varying traffic. Large antenna arrays offer an alternative: an operator can spend part of its spatial degrees of freedom (DoF) serving its own users and part suppressing interference toward others' users (nullforming), letting competing networks reuse the same band. This raises two questions: does trading DoF for interference suppression beat orthogonal partitioning, and how much coordination is needed to realize the gain? We cast inter-operator sharing as a continuum of cooperation "shades" of a single null-forming MU-MIMO precoding and interference-constrained scheduling primitive, spanning orthogonal partitioning, non-cooperative reuse, cooperative cross-operator protection, and a scheduling-aware bound exploiting foreign scheduling decisions. Evaluated on a ray-traced digital twin of a real multi-operator deployment with 27 base stations sharing a 7 GHz carrier, the cooperative shades deliver more than a 2.5$\times$ median per-user rate gain over both non-cooperative reuse and orthogonal splitting. 
Crucially, most of this gain requires only minimal inter-operator information exchange, indicating that a modest, standardizable metadata exchange, rather than tight joint processing, unlocks most of the value of shared-spectrum operation.
\end{abstract}

\section{Introduction}
Wireless demand keeps growing along many dimensions (more devices, denser deployments, higher rates, and stricter latency and reliability) yet the spectrum that can be exclusively assigned to each mobile network operator (MNO) remains scarce, especially where large bandwidth meets favorable propagation. This tension is central to 6G radio access networks (RANs), where upper-midband frequencies such as FR3 must provide both capacity and coverage, motivating a shift from rigid spectrum ownership toward more flexible sharing.

Traditional cellular spectrum management relies on exclusive, licensed, frequency-orthogonal assignments, for both nationwide licenses and spectrum granted for local or regional use, such as private or vertical deployments, where each block is reserved to one licensee and kept orthogonal to the others. Predictable and simple, this model is nonetheless inefficient \cite{voicu2018survey} under traffic heterogeneity across operators: bands cannot be reassigned when locally underutilized, and coarse frequency blocks limit fine-grained adaptation. Valuable spectrum may thus remain unused in some places and times \cite{hassan2017exclusive}, ultimately constraining the per-user rate that end-users experience and that public policies increasingly adopt as a measurable performance indicator \cite{Capone23}.

This has renewed interest in spectrum sharing and coexistence for 6G \cite{alsaedi2023spectrum, matinmikko2020spectrum}. Frameworks such as Licensed Shared Access (LSA) in Europe and Citizens Broadband Radio Service (CBRS), coordinated by Spectrum Access Systems (SAS), in the United States let users with different priorities coexist under controlled interference, but typically rely on shared databases and coarse availability information. More recently, Open RAN and AI-native RAN architectures have opened the door to more dynamic, fine-grained sharing within cellular systems \cite{testolina2024sharing}.

A particularly challenging case is sharing among MNOs in the same area. Unlike coexistence with space communication incumbents \cite{kang2024,jia2025static}, often managed via quasi-static exclusion zones, protection contours, or long-term coordination, MNO-to-MNO sharing involves networks that are simultaneously active, spatially overlapping, and independently optimized, with mobile users, fast-varying traffic, and interference evolving at scheduling and beamforming time scale. As competing operators 
have limited incentives to expose information or coordinate, inter-operator sharing demands mechanisms that are efficient, adaptive, and compatible with limited cooperation. 

Spatial processing has long mitigated interference \emph{within} a single operator's network: beamforming concentrates energy toward intended receivers and reduces leakage elsewhere. Consequently, spatial degrees of freedom (DoF) mainly serve to improve the desired link, add array gain, or support intra-operator multi-user MIMO, while inter-operator coexistence is still enforced through frequency separation. The large arrays envisioned for FR3 make it possible to use spatial DoF more deliberately as a \emph{sharing} resource: an operator can beamform toward its scheduled users while nullforming toward users of others, so transmissions from different MNOs reuse the same time-frequency resources whenever their beamforming and nullforming constraints are jointly feasible. Sharing then becomes not only a matter of who transmits on which resource, but of how concurrent transmissions are spatially arranged.

This reshapes the fundamental tradeoff: sharing the band grants extra bandwidth, but interference suppression consumes spatial DoF that could otherwise serve more users or improve link quality. The tradeoff is especially nontrivial in FR3, where, unlike millimeter-wave systems with pencil beams and high path loss, richer scattering, wider beams, and stronger inter-cell coupling make interference management a key enabler, while the same spatial richness keeps beamforming and nullforming attractive for coexistence.

We therefore study MNO-to-MNO spectrum sharing in FR3 as a challenging, underexplored 6G scenario, asking two \textit{fundamental questions}:
\begin{enumerate}
\item Can competing operators benefit from sharing the same spectrum by trading spatial DoF for interference suppression, rather than partitioning it orthogonally?
\item If so, how much inter-operator coordination is required to realize the gain?
\end{enumerate}
To answer them, we recast the seemingly discrete menu of sharing architectures as a continuum of coordination \emph{shades}, each defined by how much an operator knows about, and acts on, the users of others, and evaluate the whole continuum on a ray-traced digital twin of a real multi-operator deployment.

The remainder of the paper is organized as follows. Section~\ref{sec:systemlevel} frames null-forming-based frequency reuse at the system level and reviews related work. Section~\ref{sec:sysmodel} presents the system and signal model and states the coordination problem. Section~\ref{sec:shades} formalizes coordination as a ladder of shades, from orthogonal partitioning and non-cooperative reuse to a scheduling-aware upper bound. Sections~\ref{sec:precoding} and~\ref{sec:scheduling} detail the leakage-aware null-forming MU-MIMO precoder and the correlation-aware, interference-constrained scheduler that realize each shade. Section~\ref{sec:methodology} describes the methodology and ray-traced scenario, Section~\ref{sec:results} reports the results, and Section~\ref{sec:conclusion} concludes and outlines future work.

\section{Null-Forming as a System-Level \\Sharing Primitive}
\label{sec:systemlevel}

\begin{figure}[t]
\centering
\includegraphics[width=0.85\linewidth]{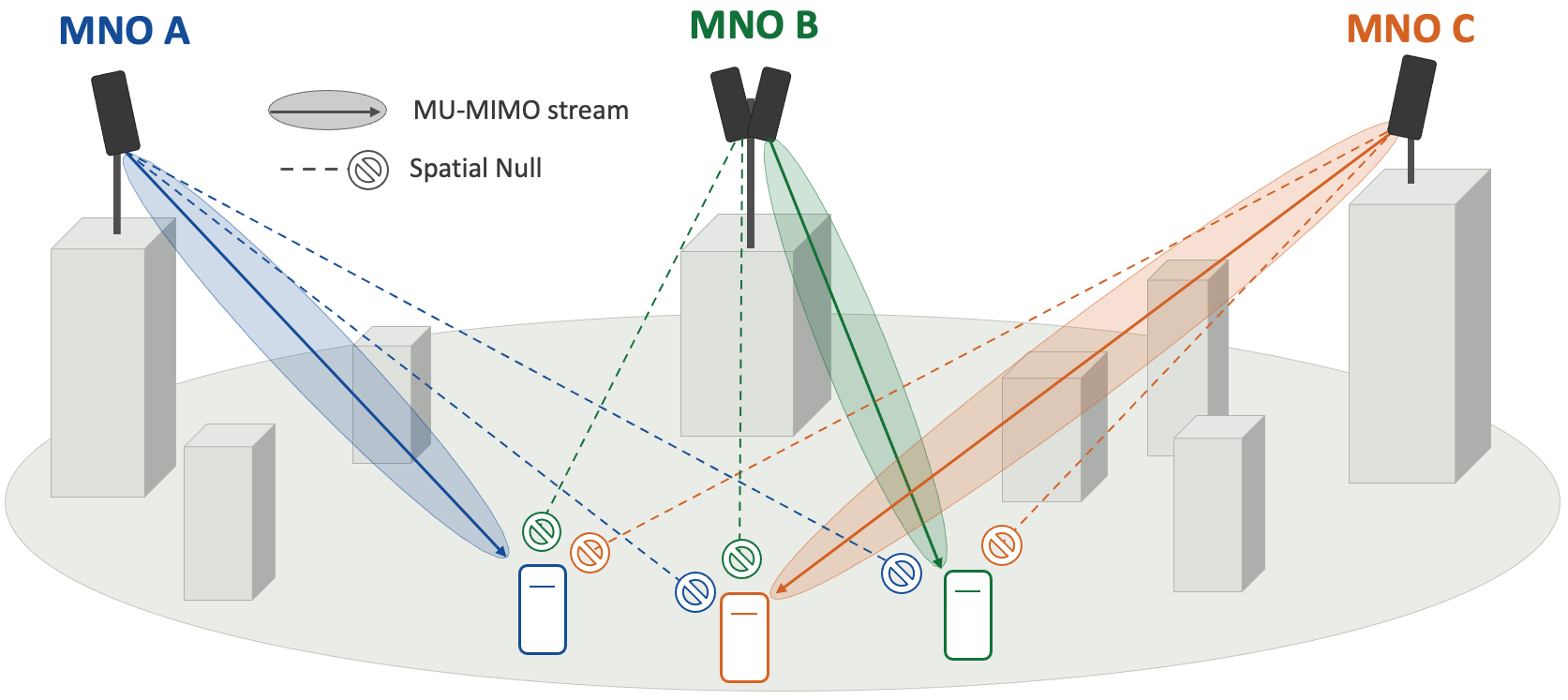}
\caption{System level envisioning: simplified scenario}
\label{fig:system}
\end{figure}

Null-forming and beamforming are usually regarded as link-level primitives: ways for a transmitter to steer beams toward its own users while placing spatial nulls toward a few interferers. Our contribution is to lift this primitive to the system level, where multiple operators reuse the same band over a shared physical environment while shaping their transmissions to protect each other's users. In this view, null-forming is not an isolated precoding choice but the technical enabler of inter-operator frequency reuse.

Realizing it, however, raises an obstacle absent from single-operator design: protection becomes more effective the more \emph{inter-operator information} is exchanged. We consider two kinds. (i) \emph{Foreign-channel acquisition}: protecting a foreign user requires some knowledge of the channel toward it; this calls for uplink sounding, a mechanism already realized intra-operator through SRS-based positioning pipelines in production O-RAN stacks \cite{MalikFlorian2025srs}. (ii) \emph{Scheduling-decision sharing}: deeper coordination further requires knowing \emph{which} foreign users are active, or even their full scheduling, i.e., on which physical resources they are served (this last case will be considered for evaluating upper limits on the performance).

Such information exchange could be implemented, for instance, through Open RAN mechanisms, with coordination orchestrated by RAN Intelligent Controllers (RICs). The definition of these interfaces and procedures is, however, out of the scope of this paper, which focuses on quantifying the potential advantages that inter-operator coordination can unlock.


\subsection{Multi-operator Spectrum Sharing}
Prior work has established the feasibility of multi-operator coexistence, but largely at two abstract levels: statistically, through stochastic-geometry interference analysis, and administratively, through spectrum-sharing policies and rules. Feasibility studies show that uncoordinated sharing is attractive precisely because, as anticipated above, narrow mmWave beams isolate transmissions almost for free \cite{gupta2016,rebato2016}; an assumption that weakens in FR3, where wider beams and richer scattering reintroduce inter-operator interference.
 
Closer to our perspective, Shokri-Ghadikolaei \emph{et al.} \cite{shokri2016} parameterize sharing by inter-operator \emph{coordination levels}, but only analytically and asymptotically, without a scheduler that realizes them. Li \emph{et al.} \cite{li2014} enforce coordination through per-link constraints exchanged between operators via a negotiation process, yet evaluate it on a single ray-traced 60~GHz corridor with two networks and no MU-MIMO precoding. Wang and Adve~\cite{wang2022} jointly optimize resource allocation and scheduling with coordinated beamforming, but restrict the study to a fixed pair of operators operating in a single cooperative mode, rather than a graded range of coordination. On the physical layer, null-forming as the underlying precoding scheme has been investigated by Tashiro \emph{et al.} \cite{tashiro2022,tashiro2025} and Kang \emph{et al.} \cite{kang2024}, and earlier in underlay cognitive radio via multi-antenna \cite{rui_sharing} and blind \cite{goldsmith_sharing} null-space transmission, but in \emph{quasi-static} heterogeneous-coexistence FR3 scenarios (aerial- and satellite-terrestrial), where interference does not evolve at the scheduling time scale.
 
To the best of our knowledge, no prior work brings the null-forming primitive together with a full coordination continuum spanning more than two operators, site-specific propagation in the upper-midband, and a scheduler whose complexity is decoupled from the resource-grid size.

\subsection{Statistical Beamforming and Group Scheduling Primitives}
Our framework rests on two building blocks, each individually studied: a leakage-aware precoder and a group-scheduling formulation. We build the precoder on \emph{long-term} beamforming (LTBF), which forms beams from slowly-varying second-order channel statistics rather than instantaneous, per-subcarrier channel state information (CSI)~\cite{caire_ltbf, rasteh2025ltbf}: a single statistically-consistent beam is robust to subcarrier-to-subcarrier variation and to channel-estimation error, and remains stable over the coherence time of the statistics. On this basis, regularized and null-forming linear precoders span a broad family, from regularized zero-forcing (RZF) and its large-system analysis under limited feedback \cite{peel2005vector,nguyen2019murfz,wagner2012rzf}, through signal-to-leakage-plus-noise-ratio (SLNR) designs \cite{sadek2007leakage}, to zero-forcing and coordinated inter-cell nulling \cite{spencer2004zero,li2010coordbf}. We introduce no new beamformer; rather, we unify this family under a single
primitive that interpolates continuously from RZF to hard cross-operator nulling, making the precoder the mechanism that realizes graded coordination.
 
On the scheduling side, grouping spatially correlated users and designing their beams are tightly coupled, yet the joint problem is combinatorial and non-convex. It is therefore almost always \emph{decoupled} in the literature, either into two sequential steps, user selection followed by beamforming \cite{he2023jointusbf,vemula2006intercell}, or by alternating between the two subproblems, holding one fixed while optimizing the other \cite{jiang2018joint,huang2016joint,kobayashi2007joint,sheemar2025joint}; both approaches yield greedy or locally optimal heuristics with problem-specific relaxations and bounded guarantees. We instead formulate an integer linear program (ILP) that integrates beamforming and user grouping through a simple correlation Gram matrix used as a spatial-separability proxy, solve it via per-operator column generation, and derive a scale-free, linear-programming (LP) ergodic relaxation whose complexity is independent of the resource-grid size.

Finally, the conclusions drawn from any such study are tied to the propagation model adopted. Prior evaluations rely on stochastic geometry \cite{gupta2016}, statistical channels \cite{rebato2016}, or a single ray-traced corridor \cite{li2014}, none of which captures a realistic multi-operator, multi-site dense-urban layout.
 
This gap defines the two technical pillars of our work: a leakage-aware LTBF precoder and a correlation-aware scheduler that jointly work according to a inter-operator coordination ladder, which instantiates orthogonal (legacy), non-cooperative, and cooperative operation as settings of the same $(\mu,\xi)$ primitive.  An upper bound on the achievable performance based on ideal cooperation is provided as well. The whole framework is evaluated on an accurate ray-traced model of a real dense-urban scenario.

\section{System Model and Coordination Setup}
\label{sec:sysmodel}

\subsection{Network and Deployment Scenario}
\label{subsec:networkscenario}
We consider a set of $N_{\mathrm{op}}$ MNOs that share a common frequency band over the same geographic area. Each operator $o$ owns a disjoint set of sites; every site is split into three sectors, and each sector is served by one base station (BS). All BSs form the set $\mathcal{B}$, partitioned across operators as $\mathcal{B}=\bigcup_{o}\mathcal{B}_o$, where $\mathcal{B}_o$ denotes the BSs belonging to operator $o$.

Each BS is equipped with a uniform planar array (UPA) of $N$ antennas and serves single-antenna user equipments (UEs). Each UE is associated with the BS of its own operator offering the strongest received power,
We denote by $\mathcal{U}_b$ the set of UEs served by BS $b$ and by $\mathcal{U}_o=\bigcup_{b\in\mathcal{B}_o}\mathcal{U}_b$ those of operator $o$. Crucially, UEs are never served by foreign BSs; they do, however, generate mutual interference whenever the shared band is reused.

Transmission follows the standard 5G New Radio (NR) time-frequency resource grid, spanning $T$ time slots and $F$ Physical Resource Blocks (PRBs). On each PRB, up to $L$ UEs can be scheduled on distinct spatial streams. 

\subsection{Signal Model}
\label{sec:signalmodel}

Consider a generic PRB and a BS $b$ serving a set $\mathcal{K}\subseteq\mathcal{U}_b$ of co-scheduled UEs, with $|\mathcal{K}|\le L$. Let $\mathbf{h}_{k}\in\mathbb{C}^{N}$ be the downlink channel from $b$ to UE $k$, and $\mathbf{w}_{k}\in\mathbb{C}^{N}$ the precoding vector assigned to it, each stream transmitted with equal power $p=P/|\mathcal{K}|$, where $P$ is the per-BS transmit power. 
The resulting post-precoding signal-to-interference-plus-noise ratio (SINR) is
\begin{equation}
  \label{eq:sinr}
  \mathrm{SINR}_{k}
  = \frac{p\,\lvert\mathbf{h}_{k}^{\mathsf{H}}\mathbf{w}_{k}\rvert^{2}}
         {\displaystyle
          \underbrace{p\!\!\sum_{j\in\mathcal{K}\setminus\{k\}}\!\!\lvert\mathbf{h}_{k}^{\mathsf{H}}\mathbf{w}_{j}\rvert^{2}}_{\text{intra-cell}}
        + \underbrace{\sum_{b'\neq b}p'\!\!\sum_{j\in\mathcal{K}_{b'}}\!\!\lvert\mathbf{h}_{k,b'}^{\mathsf{H}}\mathbf{w}_{j}\rvert^{2}}_{\text{inter-cell}}
        + \sigma^{2}}.
\end{equation}
The corresponding achievable rate is obtained by clipping the classical Shannon expression to the highest spectral efficiency among the modulation-and-coding schemes (MCSs) available in practice \cite{3gpp_speceff}: we therefore cap it with $\eta_{\max}=7.4$~bit/s/Hz, for 256-QAM
\begin{equation}
  r_k = B \,\min\!\big(\log_2(1+\mathrm{SINR}_k),\,\eta_{\max}\big).
  \label{eq:rate}
\end{equation}

\subsection{Desired and Protected Channels}
\label{subsec:desired_protected}

The interference terms in \eqref{eq:sinr} show that each BS influences two kinds of UEs. Its \emph{co-scheduled} UEs, whose channels can be stacked in $\mathbf{H}_{s}$, which are the UEs sharing the multi-user (MU) streams. Its \emph{victim} UEs, similarly stacked in $\mathbf{H}_{v}$, are those with which it must avoid interfering. This last set splits further by operator ownership:
\begin{equation}
  \label{eq:victimsplit}
  \mathbf{H}_{v} = \big[\,\mathbf{H}_{v}^{\text{same}}\;\;\mathbf{H}_{v}^{\text{cross}}\,\big],
\end{equation}
where $\mathbf{H}_{v}^{\text{same}}$ collects the victims belonging to the \emph{same} operator as the serving BS and $\mathbf{H}_{v}^{\text{cross}}$ those of \emph{other} operators. This same-versus-cross-operator partition is the structural choice on which inter-operator coordination hinges: letting each BS decide independently how strongly to suppress $\mathbf{H}_{v}^{\text{cross}}$ relative to $\mathbf{H}_{v}^{\text{same}}$ introduces a degree of freedom that future coexistence policies can build on, e.g., by imposing a cap on the maximum leakage toward other operators. These design choices are discussed further in Sec.~\ref{sec:shades}.

\subsection{Problem}
\label{sec:problemstatement}
Given the shared spectrum and the deployment described above, each operator jointly selects, at each BS, the set of UEs to co-schedule, their assigned wireless resources, and their precoders, so as to maximize a fairness utility while limiting the interference leaked onto victim UEs. Operators can perform the selection according to different inter-operator coordination approaches, as described next.


\section{The Inter-Operator Cooperation Shades}
\label{sec:shades}

\begin{table}[t]
  \centering
  \caption{The coordination shades.}
  \label{tab:shades}
  \small
  \setlength{\tabcolsep}{4pt}
  \resizebox{\linewidth}{!}{
  \begin{tabular}{@{}lccr@{}}
    \toprule
    Shade & Own-user & Foreign-user & Cross-op information \\
          & protection ($\mu$) & protection ($\xi$) & \\
    \midrule
    S0 & 0        & ---     & none (disjoint bands) \\
    S1 & tunable  & none    & none (own CSI only) \\
    S2 & tunable  & partial & full victim CSI \\
    S3 & tunable  & full    & full victim CSI \\
    S4 & tunable  & full (active only) & full CSI \emph{and} foreign schedule \\
    \bottomrule
  \end{tabular}
  }
\end{table}

Inter-operator spectrum sharing is usually presented as a discrete menu of architectures, either uncoordinated reuse, or some form of cooperation mechanism, each studied in isolation with its own model and assumptions. We argue instead that these are not distinct designs but different \emph{settings of a single primitive}, determined entirely by \emph{how much an operator knows about, and acts on, the foreign UEs}. This section makes that dependence explicit, stating the information each BS is assumed to have about foreign UEs.

\subsection{Two Weights, One Primitive}
\label{subsec:knob}

We distinguish two kinds of protected users: same-operator victims $\mathbf{H}_{v}^{\text{same}}$ and cross-operator victims $\mathbf{H}_{v}^{\text{cross}}$. The precoder we adopt (detailed in Sec.~\ref{sec:precoding}) suppresses leakage onto each set through two independent weights, $\mu$ and $\xi$, which respectively control how strongly to null same- and cross-operator victims.

Within an operator, we assume a single scheduler jointly coordinates all of that operator's BSs (see Sec.~\ref{sec:scheduling}), so each BS knows exactly which same-operator users are co-scheduled and directs its $\mu$-weighted nulls only at those \emph{scheduled} victims. Such intra-operator coordination is realistic (for instance, it can be provided by a single-DU controlling more RUs or, at a higher level, by the RAN Intelligent Controller (RIC) in an Open RAN architecture \cite{giannopoulos2026interferencegrapharxiv,reinders2026AIIMarxiv}) and it also lets us idealize intra-operator interference management, so that the study can focus on the inter-operator coordination that is our actual concern. By contrast, between different operators' BSs no mechanism currently conveys live scheduling decisions, so the cross-operator nulls must cover \emph{all potential} victims. This asymmetry between active victims that are \emph{known} within an operator versus merely \emph{knowable} across operators, is the structural assumption motivating our two-weight precoding choice.

The pair $(\mu,\xi)$ spans a continuous two-dimensional space, in which we identify four sample points corresponding to meaningful operational regimes, each defined not only by its $(\mu,\xi)$ weights but by the \emph{cross-operator information} it presumes; a fifth, idealized point (Sec.~\ref{subsec:oracle}) later serves as an upper bound. These are the shades of the coordination range.

\subsection{The Shades}
\label{subsec:shades}

\emph{S0 -- Orthogonal.} This is the current baseline: the band is partitioned into disjoint sub-bands, so operators never interfere. Cross-operator victims do not exist by construction ($\xi$ is undefined), and each BS applies a per-BS regularized precoder (typically MMSE) to its own users only. S0 needs no cross-operator information whatsoever, but pays the full price of orthogonalization: each operator is limited to a fraction of the spectrum.

\emph{S1 -- Non-cooperative.} Operators reuse the entire shared band. Each operator nulls only its \emph{own} co-scheduled victims, across all its BSs ($\mu>0,\ \xi=0$). This requires scheduling-information exchange among same-operator BSs and knowledge of own-operator CSI, but leaves interference from other operators unsuppressed.

\emph{S2 -- Cooperative (selfish).} Each BS additionally directs suppression toward cross-operator victims, but at reduced strength ($\mu>0,\ \xi=\mu/2$). It is the first shade to require cross-operator CSI. S2 captures the self-interested behavior of an operator that nulls only the \emph{bare minimum} toward competitors, retaining spatial DoF for its own beams. Such a minimum would reasonably be imposed as an interference limit that a supervising entity could standardize for coexistence; the half-weight $\xi=\mu/2$ is one representative choice on the continuous range $\xi\in[0,\mu]$. The optimization of cross-operator weights for a given fairness or protection target is itself a design question, a promising and challenging research direction that we leave for future studies.

\emph{S3 -- Cooperative (selfless).} Each BS protects cross-operator victims as strongly as its own ($\mu>0,\ \xi=\mu$), nulling every known victim regardless of operator. It requires the same victim CSI as S2 and represents the strongest coordination achievable when relying on channel knowledge alone.

\subsection{An Upper Bound: Sharing User Scheduling}
\label{subsec:oracle}

Despite its selfless cooperation, S3 does not provide the best performance, and the reason lies in the knowledge asymmetry introduced in Sec.~\ref{subsec:knob}: protecting \emph{all potential} victims inevitably wastes DoF, and thus spatial directivity, for inactive victim users, which could otherwise serve intended users.

The \emph{S4 -- Scheduling-Aware Cooperation} (SAC) shade removes this inefficiency. In addition to cross-operator CSI, it assumes inter-operator knowledge of scheduling decisions, so each BS nulls only the \emph{active} victims, i.e., co-scheduled victims. 
It therefore provides a robust upper bound on what any channel-only coordination scheme can achieve. 

\subsection{Reading the Cooperation Palette}
\label{subsec:costbenefit}

Under this view, coordination becomes a cost/benefit question with a concrete currency: deeper shades demand heavier cross-operator information in exchange for cleaner interference suppression: no signaling (S1), then increasing channel information (S2--S3), then live per-PRB schedules (S4); with the S2$\to$S3 step holding information fixed and varying only the cross-operator nulling weight $\xi$. The \textit{evaluation questions} (Sec.~\ref{sec:results}) follow directly: on the same full CSI, how much does stronger cross-operator nulling (S3) actually pay off relative to the lighter policy (S2)? And how large is the residual gap between S3 and the S4/SAC upper bound; that is, how much is live schedule sharing worth on top of channel knowledge? Since S1--S3 differ only in the weights $(\mu,\xi)$ handed to the same precoder and scheduler, the entire cooperation palette is evaluated by re-running a single pipeline across shades, with no change to the underlying machinery of Secs.~\ref{sec:precoding}--\ref{sec:scheduling}.

\section{Leakage-Aware MU-MIMO LTBF Precoding}
\label{sec:precoding}

This section develops the physical-layer primitive on which the cooperation shades of Sec. \ref{sec:shades} are built. Given a base station (BS) with $N$ transmit antennas serving its co-scheduled user set $\mathcal{K}$ while limiting interference toward a set $\mathcal{V}$ of protected (victim) users, we seek a precoder that (i) separates the desired streams, (ii) suppresses leakage onto victims, and (iii) avoids noise amplification. Consistent with the victim split of Eq. \eqref{eq:victimsplit}, leakage suppression is controlled by two independent weights: $\mu$ acts on \emph{same}-operator signals and $\xi$ on \emph{cross}-operator victims $\mathbf{H}_{v}^{\text{cross}}$. The pair $(\mu,\xi)$ is exactly the primitive whose settings define the cooperation shades of Sec. \ref{sec:shades}. Channel matrices per-PRB $f$ are written as $\mathbf{H}_s[f]\in\mathbb{C}^{|\mathcal{K}|\times N}$ (co-scheduled) and $\mathbf{H}_v[f]\in\mathbb{C}^{|\mathcal{V}|\times N}$ (victims), stacking the conjugate channel rows $\mathbf{h}_k^{\mathsf{H}}[f]$ of Sec.~\ref{sec:signalmodel}; the victim rows split as $\mathbf{H}_v= [\mathbf{H}_{v}^{\text{same}};\,\mathbf{H}_{v}^{\text{cross}}]$.

\subsection{Instantaneous Formulation (Per-PRB Baseline)}
\label{subsec:instantaneous}

When per-PRB channel state information (CSI) is reliable, the precoder $\mathbf{W}[f]\in\mathbb{C}^{N\times|\mathcal{K}|}$ minimizes on each resource a regularized cost that trades coscheduled-stream decoupling against leakage onto same- and cross-operator victims (weights $\mu,\xi\ge 0$) and noise amplification ($\lambda\ge 0$). Its closed-form solution is
\begin{equation}
\label{eq:precoder_closedform}
\mathbf{W}[f]
=\!\left(
\mathbf{H}_s^{\mathsf{H}}\mathbf{H}_s
+ \mu\,\mathbf{H}_{v}^{\text{same},\mathsf{H}}\mathbf{H}_{v}^{\text{same}}
+ \xi\,\mathbf{H}_{v}^{\text{cross},\mathsf{H}}\mathbf{H}_{v}^{\text{cross}}
+ \lambda \mathbf{I}_N
\right)^{\!-1}\!\!\mathbf{H}_s^{\mathsf{H}} ,
\end{equation}
(all matrices per-PRB $[f]$), which recovers classical strategies as limiting cases: $\mu=\xi=0$ gives conventional RZF/MMSE precoding; $\mu,\xi\to\infty$ enforces hard nulling toward same- and cross-operator victims; and $\lambda\to 0$ yields leakage-aware zero forcing. These are the per-weight limits deferred from Sec.~\ref{subsec:knob}.

\subsection{Robustness Limitation of Instantaneous Nulling}
\label{subsec:robustness}

The precoder in~\eqref{eq:precoder_closedform} suppresses leakage through \emph{coherent} cancellation, placing the columns of $\mathbf{W}[f]$ in the null space of the estimated victim channel, so the null depth is bounded by that estimate: under per-element error variance $\sigma_e^2$ the residual leakage floors at a level proportional to $\sigma_e^2\|\mathbf{w}\|^2$ regardless of the weight, and raising $\mu$ or $\xi$ past this point only amplifies noise. The floor is decisive in spectrum sharing, where the cross-operator victim channel is not obtained from dedicated downlink pilots but inferred from cross-operator sounding, whose listening-only impairments enlarge $\sigma_e^2$ well beyond the desired-link estimate. Instantaneous nulling, optimal under perfect CSI, therefore degrades sharply here, motivating the long-term (statistical) beamformer (LTBF) below.

\subsection{A Two-Weight Long-Term Beamformer (LTBF) Primitive}
\label{subsec:twoweight}

To overcome the estimation-noise floor we design on \emph{frequency-averaged spatial covariances} rather than per-PRB channels: the dominant spatial structure is essentially frequency-flat, whereas the estimation error is zero-mean and independent across PRBs, so averaging over the $F$ PRBs preserves the subspaces of interest while suppressing the error by a factor of order $F$. We define the per-user co-scheduled covariance and the aggregate same-/cross-operator victim covariances
\begin{align}
\label{eq:Rd}
\mathbf{R}_{s,k} &= \tfrac{1}{F}\!\sum_{f=1}^{F}
\mathbf{h}_{k}[f]\,\mathbf{h}_{k}^{\mathsf{H}}[f],
\qquad k \in \mathcal{K}, \\[2pt]
\label{eq:Rv}
\mathbf{R}_{v}^{\bullet} &= \tfrac{1}{F}\!\sum_{f=1}^{F}
\sum_{i\in\mathcal{V}^{\bullet}}
\mathbf{h}_{v,i}[f]\,\mathbf{h}_{v,i}^{\mathsf{H}}[f],
\quad \bullet\in\{\text{same},\text{cross}\},
\end{align}
with $\mathbf{h}_{k}[f],\mathbf{h}_{v,i}[f]\in\mathbb{C}^{N}$ the co-scheduled and victim channels of Sec.~\ref{sec:signalmodel}.

To suppress intra-BS multi-user interference jointly with leakage, we treat the co-scheduled users as further signals to null and form, for user $k$, the aggregate interference covariance
\begin{equation}
\label{eq:Rint}
\mathbf{R}_{\mathrm{int},k}
= \mu\Big(
   \underbrace{\sum_{j\in\mathcal{K}\setminus\{k\}}\mathbf{R}_{s,j}}_{\text{co-scheduled users}}
+ \underbrace{\mathbf{R}_{v}^{\text{same}}}_{\text{same-op victims}}
  \Big)
+
\xi\,\underbrace{\mathbf{R}_{v}^{\text{cross}}}_{\text{cross-op victims}}
+\;\lambda\mathbf{I}_N .
\end{equation}
This is the asymmetry on which Sec.~\ref{sec:shades} rests: the single weight $\mu$ governs \emph{all} signals the BS owns and must not radiate toward (co-scheduled users \emph{and} same-operator victims, both jointly scheduled and hence known), while the independent weight $\xi$ sets the cross-operator protection policy. The beam for user $k$ maximizes the statistical signal-to-interference-plus-leakage-plus-noise ratio,
\begin{equation}
\label{eq:slnr}
\mathbf{w}_k
= \arg\max_{\mathbf{w}}
\frac{\mathbf{w}^{\mathsf{H}}\mathbf{R}_{s,k}\,\mathbf{w}}
     {\mathbf{w}^{\mathsf{H}}\,\mathbf{R}_{\mathrm{int},k}\,\mathbf{w}},
\end{equation}
the principal generalized eigenvector of $(\mathbf{R}_{s,k},\mathbf{R}_{\mathrm{int},k})$, where the whitening factor $\mathbf{R}_{\mathrm{int},k}^{-1}$ rotates the beam away from both the co-user/same-operator and the cross-operator subspaces before capturing the desired signal. Since the eigenvector is invariant to a common scaling of $\mathbf{R}_{\mathrm{int},k}$, the direction depends on the weights only through $\xi/\mu$: a weight does not create spatial degrees of freedom (DoF), it \emph{reallocates} them. The per-user beams are collected into $\mathbf{W}=[\mathbf{w}_1,\dots,\mathbf{w}_{|\mathcal{K}|}]$, each unit-norm, then globally scaled to $\mathrm{tr}(\mathbf{W}\mathbf{W}^{\mathsf{H}})=P$ ($p=P/|\mathcal{K}|$ per stream). The resulting $\mathbf{W}$ is frequency-flat: one precoder is applied across all PRBs, in contrast to the per-PRB $\mathbf{W}[f]$ of~\eqref{eq:precoder_closedform}.

\subsection{Leakage Metric and Design Trends}
\label{subsec:metrics}

User SINR and achievable rate follow Eqs.~\eqref{eq:sinr}-\eqref{eq:rate} of Sec.~\ref{sec:signalmodel}; the complementary victim-side metric is the total leakage onto co-scheduled user $i$,
\begin{equation}
\label{eq:leak_def}
\mathcal{L}_i[f] = \sum_{k\in\mathcal{K}} |\mathbf{h}_{v,i}^{\mathsf{H}}[f]\mathbf{w}_k|^2,
\end{equation}
reported as interference-to-noise ratio $\mathrm{INR}=\mathbb{E}[\mathcal{L}_i]/\sigma^2$ (expectation over PRBs and victims).
Raising $\xi$ deepens the cross-operator null at the cost of desired-signal capture; unlike the instantaneous case, this null does not collapse under estimation error, since $\mathbf{R}_{v}^{\text{cross}}$ is denoised by averaging; the resulting trade-off is quantified in Sec.~\ref{sec:results}.

\smallskip\noindent
Finally, the same frequency-averaged covariances $\{\mathbf{R}_{s,k},
\mathbf{R}_{v}^{\bullet}\}$ that define the precoder also drive the lightweight
spatial-compatibility proxy (a normalized channel Gram) used to screen user
sets before scheduling; screening and precoding thus reason about the same
statistics. We defer it to Sec.~\ref{sec:scheduling}.


\section{Correlation-Aware Scheduling}
\label{sec:scheduling}

The precoder of Sec. \ref{sec:precoding} suppresses interference \emph{given} a set of co-scheduled users; it does not decide \emph{which} users to serve together. That's a scheduler's decision, and it matters precisely because the precoder's effectiveness degrades when co-scheduled users are spatially similar: two users with nearly parallel channels contend for the same spatial degrees of freedom, so nulling one distorts the beam toward the other. This section formulates the group scheduler that exploits the same second-order channel statistics as the precoder to select spatially compatible user groups, using an ergodic fluid approach that scales independently of the resource-grid size.

\subsection{Spatial-Compatibility Proxy}
\label{subsec:xcorr_proxy}

Evaluating the exact throughput of a candidate user group would require running the precoder of Sec. \ref{sec:precoding} for that group and every other group it competes with, which would require combinatorially many beamforming solves per PRB. We instead screen groups with a lightweight proxy built on the \emph{same} frequency-averaged covariances that define the precoder, so that the quantity the scheduler reasons about and the quantity the precoder optimizes are mutually consistent. Collecting the desired channels into the rows of $\mathbf{H}\in\mathbb{C}^{|\mathcal{K}|\times N}$ and row-normalizing to $\tilde{\mathbf{h}}_k=\mathbf{h}_k/\lVert\mathbf{h}_k\rVert$, the normalized channel Gram and its magnitude
\begin{equation}
\label{eq:gram_corr}
\mathbf{G} = \tilde{\mathbf{H}}\tilde{\mathbf{H}}^{\mathsf{H}},
\qquad
\mathbf{C} = |\mathbf{G}|,
\qquad
\rho_{kj} = C_{kj}\in[0,1],
\end{equation}
give a pairwise spatial-correlation matrix: $\rho_{kj}\!\to\!0$ for orthogonal (separable) users and $\rho_{kj}\!\to\!1$ for aligned ones. A user group is a good scheduling candidate when its off-diagonal correlations are small.

We stress that low pairwise correlation is \emph{necessary but not sufficient} for good separability: $\mathbf{C}$ captures pairwise alignment but not the joint conditioning of three or more channels, and it ignores the victim subspace. It is deliberately a screening proxy, not a rate predictor; its role is to prune the pattern search cheaply, while realized rates are always evaluated with the true precoder and true channels (Sec. \ref{sec:results}).

\subsection{Exchangeable Resources and the Representative Resource}
\label{subsec:lp_ergodic}

Before committing to decision variables we exploit a structural property of the resource grid that removes the need to index by $(t,f)$ at all. Under the statistical LTBF description of Sec. \ref{sec:precoding} the precoder is frequency-flat and each UE is described by covariance statistic that's not PRB-specific, so the blocks of the $\mathcal{T}\times\mathcal{F}$ grid are \emph{statistically exchangeable}: no scheduling decision has any reason to prefer one block over another. Any optimal allocation is therefore invariant to \emph{which} resource a group occupies. The intuition is that we can solve the scheduling problem only on one \emph{representative resource}, and then generalize the outcome to scale it freely across the resource-grid.

A scheduling \emph{pattern} $p$ is a feasible group of co-scheduled users on that representative resource, encoded by the participation indicator $a_{k,p}\in\{0,1\}$, and $\mathcal{P}$ collects all feasible patterns (a pattern scheduling at most $L$ users, one layer each, per BS). The scheduler chooses a distribution $\{\lambda_p\}_{p\in\mathcal{P}}$ over patterns,
\begin{equation}
\label{eq:pattern_dist}
\sum_{p\in\mathcal{P}}\lambda_p = 1,\qquad \lambda_p\ge 0 ,
\end{equation}
that should be read as a \emph{resource-sharing} distribution: pattern $p$ is played on a fraction $\lambda_p$ of the grid. The full-grid schedule follows by replication, playing each pattern on a proportional amount of blocks. Because the resources are exchangeable, this reconstruction realizes the chosen $\{\lambda_p\}$ arbitrarily closely as $|\mathcal{T}|\,|\mathcal{F}|$ grows, the integrality gap vanishes in the limit.
The formulation below is therefore not a relaxation of a grid-indexed ILP, but its ergodic (fluid) limit; we solve it once, at a size independent of $|\mathcal{T}|\,|\mathcal{F}|$, and scale it grid-wide, rather than building a grid-coupled integer schedule and analyzing a particular ILP solution.

\subsection{Correlation-Aware Scheduling Problem}
\label{subsec:service_metric}

Classical schedulers apply their fairness objective to the number of allocated resource blocks or to the raw throughput. Neither is appropriate here: raw resource count ignores spatial quality, and exact throughput is the expensive quantity we are trying to avoid computing inside the optimization. We therefore weight each allocation by a spatial-quality factor derived from the proxy \eqref{eq:gram_corr}. Within a pattern $p$, the degradation seen by user $k$ is the largest correlation to any of its co-members,


which yields the pattern quality and the per-user service
\begin{equation}
\label{eq:pattern_quality}
Q_{k,p} = a_{k,p}\left(1-\max_{j\in p,\,j\neq k}\ \rho_{kj}\right),
\qquad
S_k = \sum_{p\in\mathcal{P}} Q_{k,p}\,\lambda_p .
\end{equation}
$Q_{k,p}$ approaches 1 when $k$'s co-members are near-orthogonal and shrinks as poorly separable users are added, so $S_k$ is the correlation-weighted service delivered to $k$ under the pattern distribution. The scheduler then solves
\begin{subequations}
\label{eq:master}
\begin{align}
\max_{\{\beta,\lambda_p\}}\ \ & \beta
+ \epsilon\sum_{k \in \mathcal{K}} S_k
\label{eq:master_obj}\\
\text{s.t.}\ \
& \textstyle\sum_{p\in\mathcal{P}}\lambda_p = 1 ,
\label{eq:master_select}\\
& \beta \leq S_k
&& \forall k\in\mathcal{K} ,
\label{eq:master_service}\\
& \beta,\, \lambda_p \ge 0
&& \forall p\in\mathcal{P} .
\label{eq:master_nonneg}
\end{align}
\end{subequations}
%
This objective maximizes the minimum per-user service and, among solutions attaining the same minimum service, the average per-user service across users ($\epsilon \sim 10^{-4}$). This fairness is achieved in terms of spatial compatibility rather than physical-layer throughput. Therefore, although spatial correlation is a desirable proxy for the effectiveness of multi-user transmissions, it does not capture channel path loss or interference power. Consequently, a positive value of $S_k$ does not necessarily guarantee a minimum throughput for user $k$, particularly for users experiencing unfavorable cell-edge conditions. The actual rates associated with the selected schedule must therefore be evaluated using the precoder and the instantaneous channel realizations, as described in Sec.~\ref{sec:results}.

\subsection{Per-Operator Column Generation}
\label{subsec:colgen}

Since $\mathcal{P}$ is exponentially large, \eqref{eq:master} is solved by means of Column Generation (CG) \cite{bjorklund2003resource,capone2010solving}. The solution starts with a restricted set of patterns $\bar{\mathcal{P}}\subseteq\mathcal{P}$, which is expanded by CG pricing. Associating the dual $\pi_k\ge 0$ with the service constraint \eqref{eq:master_service}, and $\sigma\in\mathbb{R}$ with the selection constraint \eqref{eq:master_select}, a candidate pattern $p$ has reduced cost $\bar c_p=\sum_{k}\pi_k Q_{k,p}-\sigma$ and is worth adding to the master when $\bar c_p>0$. The pricing step searches the whole pattern space for the most profitable column:
\begin{subequations}
\label{eq:pricing}
\begin{align}
p^\star \in \arg \max_{a,\gamma}\ \ & \sum_{k\in\mathcal{K}}\pi_k\,a_k\,(1-\gamma_k)
\label{eq:pricing_obj}\\
\text{s.t.}\ \
& \textstyle\sum_{k\in\mathcal{K}_b} a_k\le L
&& \forall b
\label{eq:pricing_layers}\\
& \gamma_k \ge \rho_{kj}\,a_j
&& \forall k,\,j\neq k
\label{eq:pricing_corr}\\
& a_k\in\{0,1\},\quad 0\le\gamma_k\le 1 ,
\label{eq:pricing_dom}
\end{align}
\end{subequations}
where $a_k$ marks membership of user $k$ in the candidate pattern, \eqref{eq:pricing_layers} enforces the per-BS layer budget, and \eqref{eq:pricing_corr} sets the intra-pattern degradation $\gamma_k=\max_{j\neq k}\rho_{kj}a_j$. Column $p^\star$ is added if its reduced cost $\bar{c}_p$ is positive; otherwise the current master solution is optimal for the full LP. Intuitively, pricing hunts for a group that is both valuable under the fairness duals $\pi_k$ and spatially compatible (small $\gamma_k$).

\subsection{Per-operator decomposition}
Crucially,  \eqref{eq:master}-\eqref{eq:pricing} contain no cross-operator coupling: each operator's users, layer budgets, and correlations are its own, so column generation runs \emph{independently per operator}. The optimization thus emits, for each operator independently, a distribution $\{\lambda_p\}$ over spatial patterns. Operators couple only through the precoding stage of Sec.~\ref{sec:precoding}, where the $(\mu,\xi)$ weights set how strongly each BS nulls the others.

Indeed, under the \textit{S2} and \textit{S3} cooperation schemes, all foreign users are treated as protected users, regardless of whether they are simultaneously scheduled by their respective operators. This removes the need to coordinate scheduling decisions across operators. Nevertheless, the rates ultimately experienced by the users still depend on the joint scheduling realization across operators, since it determines which users are simultaneously active and hence the actual interference and precoding conditions. 
This effect is evaluated in terms of per-UE rate, SINR, and interference statistics as described with the evaluation procedure in Sec. \ref{subsec:eval_protocol}.



\section{Evaluation Methodology}
\label{sec:methodology}

We evaluate the framework on a Sionna-RT \cite{Hoydis2023sionna} replica of a real dense-urban site, so that the channel matrices that drive both the precoder (Sec. \ref{sec:precoding}) and the scheduler (Sec. \ref{sec:scheduling}) reflect site-specific geometry rather than a stochastic model. The full simulation configuration is collected in Table \ref{tab:sim_params}; the remainder of this section describes the scenario, channel generation, and evaluation protocol.

\begin{table}[t]
  \centering
  \caption{System parameters.}
  \label{tab:sim_params}
  \small
  \setlength{\tabcolsep}{4pt}
  \begin{tabular}{@{}ll@{}}
    \toprule
    Parameter & Value \\
    \midrule
    Scenario            & University campus neighborhood\\
    Ray tracer          & Sionna RT, ITU concrete material \\
    Sites / sectors     & 9 sites $\times$ 3 sectors (27 BSs) \\
    Operators           & 3 (3 sites each) \\
    Sector azimuth / downtilt & $0^\circ-120^\circ-240^\circ$ / $10^\circ$ \\
    Carrier $f_c$       & 7~GHz (FR3) \\
    Numerology / SCS    & $\mu=1$ / 30~kHz \\
    Bandwidth           & 273 PRBs ($\approx$100~MHz) \\
    Covariance subsampling & 32 PRBs, uniformly spaced \\
    BS array            & $16\times16$ UPA, $N=256$ \\
    Tx power            & 30~dBm \\
    Noise PSD / figure  & $-174$~dBm/Hz / 7~dB \\
    UL pilot power / SRS reps & 200~mW / 4 \\
    BSs load            & 10 UEs/BS \\
    MC realizations     & 20 \\
    MU-MIMO layers $L$  & 2, 4, 8  \\
    Weight sweep $\mu$  & $\{0.1, 0.5, 1, 3, 10, 50\}$ \\
    Spectral efficiency cap            & 7.4~bit/s/Hz (256-QAM) \\
    \bottomrule
  \end{tabular}
\end{table}

\subsection{Scenario and Channels}
\label{subsec:scenario}

\begin{figure}[t]    
\centering
    \includegraphics[width=0.95\linewidth]{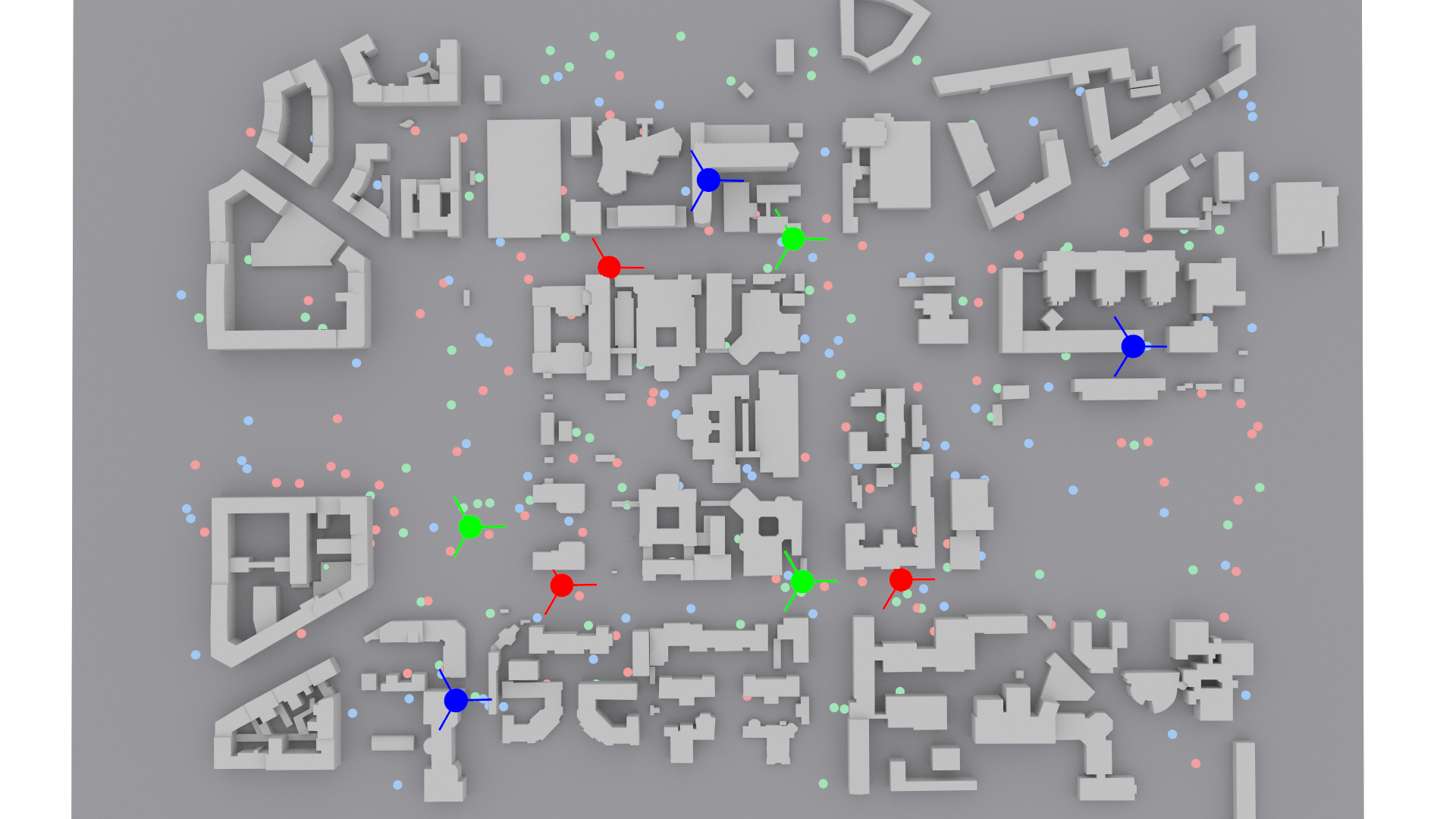}
    \label{fig:render}
  \caption{3D render of the ray-traced scenario}
  \label{fig:scenario}
\end{figure}

The scenario in Fig. \ref{fig:scenario} is a ray-traced mesh of our university campus, with real buildings' footprints and ITU concrete properties. The three-sector sites use the real positions of commercial operators in the area, grouped into three operators sharing one FR3 carrier across the full band. UEs are dropped uniformly and associated to the strongest sector; the load is fixed at $10$ UEs per sector/BS so schemes are compared at a common operating point. The long-term covariances $\mathbf{R}_{s,k}$ and $\mathbf{R}_v^{\bullet}$ of Sec. \ref{sec:precoding} are estimated over equally spaced PRBs across the band, this sampling procedure preserves the frequency diversity of the channel statistics while limiting the computational cost. To make the estimation error realistic, we derive the CSI noise from the actual SRS setup (power and repetitions) rather than assuming a generic noise floor; the resulting estimate $\mathbf{\hat{h}}$, drives precoder design. Metrics are pooled over a set of 20 independent realizations.

\subsection{Schemes and Sweeps}
\label{subsec:schemes}

All schemes share the same $(\mu,\xi)$ precoder from Sec. \ref{sec:precoding}, and per-operator scheduler from Sec. \ref{sec:scheduling}, differing only in the cooperation shade of Sec. \ref{sec:shades}: orthogonal (S0), non-cooperative (S1), cooperative (S2-S3), and the scheduling-aware S4 that upper-bounds any channel-only gain (Table \ref{tab:shades}). For each shade we sweep weight $\mu$ over the range in Table \ref{tab:sim_params}, spanning from regularized zero-forcing to hard nulling, with $\xi$ following the shade's $\xi/\mu$ tie.

\subsection{Metric Evaluation}
\label{subsec:eval_protocol}

Although pattern distributions $\{\lambda_p^{(o)}\}$ are obtained independently by operators, UEs' metrics such as SINR and inter-cell INR depend on other operators' scheduled patterns, since those determine the residual leakage it receives. We therefore evaluate metrics on the \emph{joint} scheduling state. As the operators schedule independently each pattern in the grid, fractions $\lambda_p$ play the role of stationary scheduling probabilities, therefore we define as a joint state a tuple of one pattern per operator with product probabilities
\begin{equation}
\label{eq:joint_state}
\Pr\{(p_1,\dots,p_O)\} = \prod_{o=1}^{O}\lambda_{p_o}^{(o)} .
\end{equation}
For each joint state we form the combined active user set, compute the $(\mu,\xi)$ precoder of Sec. \ref{sec:precoding} on the frequency-flat statistical channel, and evaluate per-UE SINR, SNR, and the inter-cell INR against the \emph{true} ray-traced channels of Sec. \ref{subsec:scenario}; the per-UE rate follows from \eqref{eq:rate}.
Each joint-state sample contributes to each per-UE distribution weighted by the joint probability \eqref{eq:joint_state}, then a CDF is taken over all UEs.

The cardinality of the joint state space grows rapidly with the number of patterns. To limit the evaluation cost of performance metrics, we sort the joint states by $\lambda_p$ and retain, after renormalization, the most probable ones. Section~\ref{subsec:res_validation} validates this truncation by examining the convergence of the reported metrics as the number of retained states increases.
This approach produces metrics that are \emph{ergodic} quantities: an expectation over the scheduler's own pattern distribution, evaluated on site-specific channels, not the merely outcome of one arbitrary time-frequency assignment.

\section{Results}
\label{sec:results}

We report per-UE experienced rate,
ergodic SINR,
SNR, and \emph{inter-cell} INR. Coexistence fairness is summarized by the mean, the median, and $95$th percentile per-UE. All curves follow the joint-state evaluation procedure of Sec. \ref{subsec:eval_protocol}, computed considering true ray-traced channels.

\subsection{Validating the Ergodic Metric Evaluation}
\label{subsec:res_validation}

\begin{figure}[t]
    \centering
    \includegraphics[width=0.85\columnwidth]{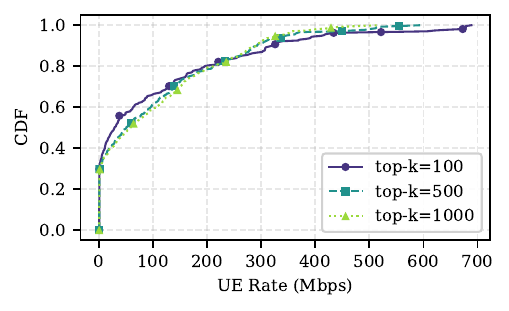}
  \caption{Representativeness of most frequent joint patterns}
  \label{fig:topk}
\end{figure}

The UE rate is computed over the top joint pattern states, ranked by their probability from \eqref{eq:joint_state}, as explained in Sec.~\ref{subsec:eval_protocol}.
Fig. \ref{fig:topk} evaluates the UE rate CDF over the top $100$, $500$, and $1000$ joint states. While the 100-state curve still presents some convergence issues, $500$- and $1000$-state curves are almost overlapped: considering the top $500$ joint states essentially captures the whole system behavior, without substantially lacking generality. We therefore fix the cap at $500$ joint states for all subsequent results, which caps evaluation cost with no measurable loss of accuracy.




\begin{figure*}[t]
  \centering
  \includegraphics[width=\linewidth]{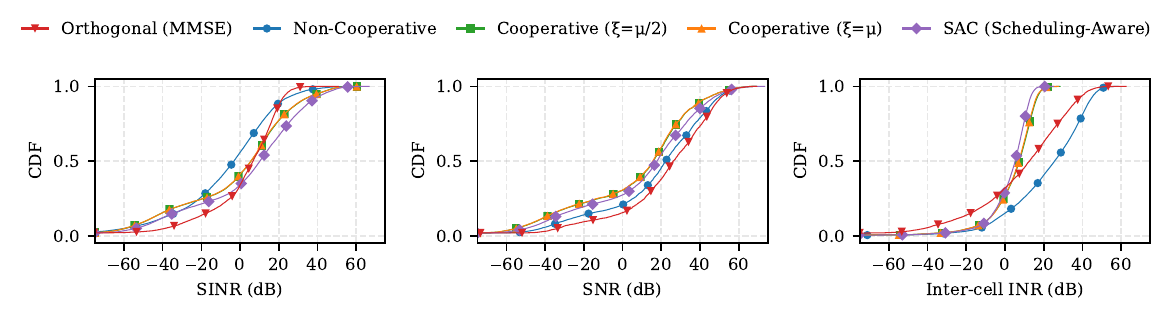}
\end{figure*}

\begin{figure*}[t]
\vspace{-0.8cm}
  \centering
  \resizebox{1\linewidth}{!}{
  \begin{subfigure}{0.40\linewidth}
    \centering
    \caption{}
    \label{fig:sinr}
  \end{subfigure}
  \hfill
  \begin{subfigure}{0.30\linewidth}
    \caption{}
    \label{fig:snr}
  \end{subfigure}
    \hfill
  \begin{subfigure}{0.35\linewidth}
    \centering
    \caption{}
    \label{fig:inr}
  \end{subfigure}}
  \caption{Main metrics CDFs, best $\mu$ per approach}
  \label{fig:metrics}
\end{figure*}

\begin{figure}[t]
\vspace{-0.5mm}
  \centering
  \includegraphics[width=0.85\linewidth]{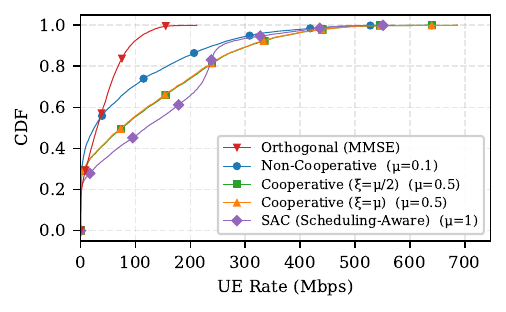}
  \caption{Per-UE rate CDF at each shade's best $\mu$}
  \label{fig:rate_cdf}
\end{figure}

\subsection{Cooperation Policies Comparison}
\label{subsec:res_ladder}
For each policy shade, we empirically select nulling weight $\mu$ that maximizes the mean UE rate; the resulting $\mu^\star$ is $0.1$ for non-cooperative reuse (S1), $0.5$ both light- and heavy- nulling cooperative (S2-S3), and $1$ for S4 SAC, while orthogonal (S0) uses no cross-operator weight. Fig. \ref{fig:rate_cdf} shows the per-UE rate CDF and Table \ref{tab:rate_cdf} the corresponding percentiles; Figg. \ref{fig:sinr}-\ref{fig:inr} report related SINR, SNR, and inter-cell INR values. 

The SNR plot (Fig.\ref{fig:snr}) focuses on the target signal gain and shows this order from worst to best received signal strength: cooperative shades (S2--S3), SAC (S4), non-cooperative reuse (S1), and orthogonal (S0); the order indeed reflects the number of nulls placed by each shade and thus the gain sacrificed in main beams directivity. The SINR panel (Fig. \ref{fig:sinr}) shows this ordering reverse once interference is accounted for: non-cooperative reuse keeps its high SNR but its SINR tail collapses under unsuppressed cross-operator leakage, clear evidence that raw signal strength isn't the right currency when operators share the same band. The shades that spend DoF on nulling convert ilower SNR into markedly higher SINR, confirming that trading array gain for interference suppression is, empirically, a favorable exchange rather than a wash. 

Per-UE rate plots in Fig.~\ref{fig:rate_cdf} and Table \ref{tab:rate_cdf} show a clear separation between the cooperation shades and the two baselines. Orthogonal S0 pays for its interference-free bands with the lowest rates everywhere (median $\approx\!30$ Mbps, 95th percentile capped near $112$~Mbps): stiffly splitting the spectrum wastes multiplexing capacity. Non-cooperative reuse recovers the high-rate tail but leaves the worst cell-edge, because it ignores the cross-operator leakage, visible as its far-right inter-cell INR distribution (median $\approx\!26$ dB in Fig. \ref{fig:inr}). Introducing cross-operator protection (S2/S3) reduces that inter-cell INR by almost $20$ dB at the median and lifts every rate percentile: the cooperative shades (S2-S3) raise the median UE rate to $\approx\!75$ Mbps, roughly $3\times$ the non-cooperative median. The S4 SAC, which additionally exploits the foreign users scheduling, is best in terms of median and mean statistics, respectively $\approx\!119$ and $\approx\!130$ Mbps. About $10\%$ above S3 in the mean and over $50\%$ in the median. Note that these percentages upper bound the performance gains that any possible channel-only cooperation scheme can reach.

A closer look at the two cooperative shades shows them nearly indistinguishable: half-weight nulling (S2, $\xi=\mu/2$) matches or marginally exceeds full nulling (S3, $\xi=\mu$) at every rate percentile (Table \ref{tab:rate_cdf}), even though S3 drives the inter-cell INR lower. This is due to the same degrees-of-freedom tradeoff seen in the SNR panel: once the residual leakage already sits below the noise-plus-intra-cell floor, deepening the null buys negligible SINR while consuming spatial DoF that S2 instead retains for the target beam. 

\subsection{Discussion of Coordination Value}
The results of this section answer the questions posed in Sec.~\ref{subsec:costbenefit} directly. First, spatial DoF is worth trading for interference suppression: any cooperative shade beats orthogonal and non-cooperative baselines by more than $2.5\times$. Second, that gain requires little coordination: S2's light nulling matches S3's full nulling, so the marginal value of $\xi=\mu$ over $\xi=\mu/2$ is negligible once leakage already sits below the noise floor. Third, the residual gap to S4/SAC, the scheduling-aware upper bound, quantifies what live schedule sharing is worth on top of channel knowledge alone: about 10\% in mean rate and over 50\% at the median, an upper bound on any channel-only coordination scheme. We further verified that this gap is modulated by the number of MU-MIMO layers $L$: repeating the full evaluation at $L=2,4,8$ (fixed 10-UE/BS load, results omitted for brevity) shows the SAC advantage shrinking monotonically as $L$ grows, since with more layers every shade can already serve a larger share of active users regardless of coordination, leaving less room for foreign-schedule knowledge to help.



\begin{table}[t]
  \centering
  \caption{UE rate (Mbps) percentiles at the best $\mu$.}
  \label{tab:rate_cdf}
  \resizebox{\linewidth}{!}{
  \begin{tabular}{lllrrrr}
    \toprule
    Approach & $\mu^{\star}$ & median & p95 & mean \\
    \midrule
    S0 - Orthogonal (MMSE) & -- & 30.13 & 111.89 & 38.38 \\
    S1 - Non-Cooperative & 0.1 & 24.46 & 309.01 & 78.12\\
    S2 - Cooperative ($\xi=\mu/2$) & 0.5  & 75.56 & 371.63 & 118.15\\
    S3 - Cooperative ($\xi=\mu  $) & 0.5 & 74.81 & 371.02 & 116.95\\
    S4 - SAC (Scheduling-Aware) & 1 & 119.05 & 335.91 & 130.65\\
    \bottomrule
  \end{tabular}
  }
\end{table}
\section{Conclusion and Future Directions}
\label{sec:conclusion}

We presented a spectrum-sharing framework built on a two-parameter $(\mu,\xi)$ precoding primitive that jointly controls intra-operator multiplexing and inter-operator interference suppression and, combined with correlation-aware, scale-free scheduling, expresses cooperation \emph{shades} through the information exchanged across operators. On a ray-traced reconstruction of a real multi-operator deployment, cooperative sharing yields over a $2.5\times$ median UE-rate gain against both orthogonal partitioning, which wastes multiplexing, and non-cooperative reuse, which incurs cross-operator interference. Most of this gain requires limited coordination: channel-derived victim information nearly matches far more informed schemes.

Several directions remain open. First, the ratio $\xi/\mu$ could adapt to network conditions, such as load, array size, victim count, and measured leakage, rather than being fixed per shade.
Second, the residual gap to scheduling-aware coordination (SAC) motivates lightweight scheduling-information exchange: existing O-RAN interfaces and SRS-related metadata can convey the victim-channel covariance our precoder needs without exposing raw CSI.

More broadly, shared-spectrum gains require effective physical-layer mechanisms, interoperable standards, and viable inter-operator agreements. Since a modest, standardizable metadata exchange already captures most of the gain, the open challenges are coordination interfaces, incentives, and governance mechanisms, including market-based ones.


\bibliography{references_ha}{}
\bibliographystyle{IEEEtran}

\vfill

\end{document}